\documentstyle[12pt]{article}

\newcommand{\eq}{\begin{equation}}
\newcommand{\en}{\end{equation}}

\def\spinst#1#2{{#1\brack#2}}

\newcommand{\NP}[1]{Nucl.\ Phys.\ {\bf #1}}
\newcommand{\PL}[1]{Phys.\ Lett.\ {\bf #1}}

\newcommand{\PR}[1]{Phys.\ Rev.\ {\bf #1}}
\newcommand{\PRL}[1]{Phys.\ Rev.\ Lett.\ {\bf #1}}

\begin{document}

\hskip 12cm \vbox{\hbox{DFTT/SM 37}\hbox{August 1991}}
\vskip 0.4cm
\centerline{\bf CRITICAL BOUNDARY CONDITIONS }
\centerline{\bf FOR THE EFFECTIVE STRING}
\vskip 1.3cm
\centerline{ M. Caselle$^{a,\diamond}$ , R. Fiore$^b$ ,
F. Gliozzi$^a$, P. Provero$^c$ and S. Vinti$^a$}
\vskip .6cm
\centerline{\sl $^a $ Dipartimento di Fisica
Teorica dell'Universit\`a di Torino}
\centerline{\sl Istituto Nazionale di Fisica Nucleare,Sezione di Torino}
\centerline{\sl via P.Giuria 1, I-10125 Turin,Italy}
\vskip .2cm
\centerline{\sl $^b$ Dipartimento di Fisica, Universit\`a della Calabria}
\centerline{\sl Istituto Nazionale di Fisica Nucleare, Gruppo collegato di
Cosenza}
\centerline{\sl Arcavacata di Rende, I-87036 Cosenza, Italy}
\vskip .2cm
\centerline{\sl $^c$ Dipartimento di Fisica, Universit\`a di Genova}
\centerline{\sl Istituto Nazionale di Fisica Nucleare,Sezione di Genova}
\centerline{\sl V. Dodecaneso 33, I-16146 Genova, Italy}
\vskip 1.cm

\begin{abstract}
Gauge systems in the confining phase induce constraints at the boundaries of
the effective string which rule out the ordinary bosonic string even
with short distance modifications. Allowing topological excitations,
corresponding to winding around the colour flux tube, produces at the quantum
level a universal free fermion string with a boundary phase $\nu={1\over 4}$.
This coincides with a model proposed some time ago in order to fit Monte Carlo
data of 3D and 4D Lattice gauge systems better. A universal value of the
thickness of the colour flux tube is predicted.
\end{abstract}
\vskip 1.5cm
\hrule
\vskip.2cm
\noindent

\noindent
$^{\ast}${\it Work supported in part by Ministero dell'Universit\`a e della
Ricerca Scientifica e Tecnologica}

\hbox{\vbox{\hbox{$^{\diamond}${\it email address:}}\hbox{}}
 \vbox{\hbox{ Decnet=(39163::CASELLE)}
\hbox{ Bitnet=(CASELLE@TORINO.INFN.IT)}}}
\vfill
\eject

\newpage
\setcounter{page}{1}

The long distance dynamics of any gauge field theory in the confining
phase
is most likely described by an effective string theory
{}~\cite{{NO},{PO}}.
A rather general way to formulate this hypothesis is to assume that the
 vacuum expectation values of gauge invariant quantities involving
large loops are expressible in terms of a two-dimensional conformal
field theory on a Riemann
surface with these loops as boundaries.

\vskip.2cm
In particular, in a D-dimensional gauge system, at finite temperature
$T=1/L$ or, equivalently, on a lattice of size
$L\ \times\ \infty^{D-1}$,
the correlation function of two Polyakov lines $P(x)$ parallel to the
periodic time axis at a distance $R$ is given by
\eq
\langle P(x)P^\dagger(x+R) \rangle = {\rm e}^{-F(R,L)}~~~~,
\label{polya}
\en
\noindent
where $F(R,L)$ is the free energy of the effective conformal theory on the
cylinder of height $R$ bounded by the two Polyakov lines of length $L$.
Similarly, the vacuum expectation value of a large, rectangular Wilson loop
is related to the free energy of a rectangle.

\vskip.2cm
In conformal field theory a central role is played by the conformal anomaly
$c$, which measures the response of the dynamical system to curving
of the surface and
also controls the finite size scaling .

\vskip.2cm
In eq. (\ref {polya}) we can consider two different finite size configurations.
In the large $R$ limit eq. (\ref{polya}) yields the free energy of an
infinitely
long cylinder having the asymptotic expansion ~\cite{cardy}
\eq
F(R,L)/R=\sigma L - {{c' }\over 6} {\pi \over L} + O ({1\over {L^2}})
\label{frasy}
\en
where $\sigma$ is the string tension, $c'=c-24 h$ is the effective
conformal
anomaly~\cite{itz} and  $h$ is the lowest conformal weight of the
states which can propagate
along the cylinder. If the theory is unitary, the ground state has $h=0$.
The term $-{\pi c'/ 6L}$ can be viewed as a universal correction
of the string tension due to the finite temperature $T={1/ L}$, i.e.
{}~\cite{pisarski}
$\sigma(T)=\sigma-\pi c'T^2/6+O(T^3)$.

\vskip.2cm
If we take the other limit $L\to\infty$ keeping $R$ fixed, the surface
of the conformal system becomes the world sheet of an open string, i.e.
an infinitely long strip of width $R$. The free energy has now a slightly
different asymptotic expansion ~\cite {cardy2} depending on the
boundary conditions
on either side of the strip. Labelling this pair of conditions
with $\alpha$ and $\beta$ we have
\eq
V(R)=\lim_{L\rightarrow\infty}{F(R,L) \over L} =
\sigma R + k_{\alpha \beta}
-({c\over 24}-h_{\alpha \beta}){\pi \over R}+O({1 \over R^2})
\label{frasy2}
\en
where $V(R)$ is the interquark potential,
$k_{\alpha\beta}$ a non-universal
constant and $h_{\alpha\beta}$ the lowest conformal weight of the spectrum
of string states compatible with the boundary conditions $\alpha \beta$;
in the particular case in which $\alpha=\beta$ the lowest state is the ground
state with $h_{\alpha\alpha}=0$. The universal term of eq. (\ref{frasy})
and (\ref{frasy2}) is a two-dimensional analog of the Casimir effect and
generalizes the evaluation of the vacuum energy of a free string
{}~\cite{{brink},{luscher}} to an arbitrary conformal theory .

\vskip.2cm
{}From the foregoing discussion, we see that the effective conformal theory
of a gauge system is completely specified not only by the conformal
anomaly $c$ and the spectrum of the physical states, but also by a
specific choice of the boundary conditions $\alpha$ and $\beta$ on
either side of the strip. Gauge theory poses two important constraints
on the latter two.

\vskip.2cm
Suppose deforming the strip in such a way that the two sides overlap:
if the two boundary conditions $\alpha$, $\beta$ were compatible, we
should get the conformal theory on a cylinder. On the other hand, from
the point of view of the gauge theory, when a quark line overlaps an
antiquark one, the free energy vanishes identically; this is possible
only if $\alpha$ and$\beta$ are incompatible, i.e. the set of states
obeying both $\alpha$ and $\beta$ on
the same side is empty. We can represent this situation symbolically by
\eq
\alpha\cap\beta=\emptyset~~~,
\label{empty}
\en
which implies, in particular
\eq
h_{\alpha\beta}>0~~~,
\label{h}
\en

because the ground state cannot propagate if $\alpha\not=\beta $
{}~\cite{cardy2}.

\vskip.2cm
The other constraint comes simply from the fact that the two sides of the
string
are intrinsically indistinguishable, so there must exist a $Z_2$
automorphism
of the theory which transposes $\alpha$ and $\beta$
\eq
Z_2:\alpha\leftrightarrow\beta\ \ \ F(L,R)\leftrightarrow F(L,R)
\label{z2}
\en

As an example of a conformal theory obeying (\ref{empty} - \ref{z2}), consider
the critical Ising model on an infinite strip in which in one side the boundary
spins are all in the $s=1$ state ($\alpha$) and in the other side they are all
in the $s=-1$ state ($\beta$). Clearly $\alpha\cap\beta=\emptyset$ and the
$Z_2$ automorphism is obviously $s\leftrightarrow -s$; in this example
$c={1\over2}$ and $h_{\alpha\beta}={1 \over 2}$ ~\cite{cardy2}.

\vskip.2cm
What is the value of $c$ and $h_{\alpha\beta}$ of the effective conformal
theory? The simplest assumption , suggested by the well established area law
of the large Wilson loops, is that the effective theory is described by a
Nambu-
Goto string with fixed boundaries. The only relevant dynamical variables
are the transverse displacements $X_i\ (i=1,\dots,D-2)$ of the string. In the
light-cone gauge, the Nambu-Goto action can be written as $D-2$ copies of
a free bosonic field with action
\eq
S_0={\sigma \over 2}\int d\tau d\varsigma \,\partial^a  X_i
\partial_a X^i~,~~a=1,2
\label {action}
\en
This gives $c'=c$ and
\eq
c'=D-2
\label{c}
\en
Alternatively, in the covariant gauge,
the Nambu-Goto action can be written as the sum of $D$
copies of free bosons with $c'_{matter}=c_{matter}=D$ and the non-unitary
system of Fadeev-Popov ghosts $b$, $c$ of conformal weight 2 and -1,
respectively, with $c_{ghost}=-26$, $c'_{ghost}=c_{ghost}+24=-2$, giving again
$c'=c'_{matter}+c'_{ghost}=D-2$.

\vskip.2cm
There are, however, notorious troubles in applying the bosonic string
outside the critical dimension of 26: depending on the quantization
method, one finds either the breaking of the Lorentz invariance or the
appearance of longitudinal string oscillations. Olesen has shown
{}~\cite{olesen} that these troubles asymptotically disappear at large
distances.

\vskip.2cm
If one requires a
consistent string model at shorter distances, one can add to the
Nambu-Goto
action with $D<26$ suitable interaction terms ~\cite{polchi} such that
$c_{matter}
=26$ in order to preserve the transversality and the Lorentz invariance.
It turns out that the resulting conformal theory has operators of
negative weight even in the matter sector, then ~\cite{itz}
$c'_{matter}\not=c_{matter}$; nevertheless one still has ~\cite{polchi}
 $c'=D-2$. This result is not very
surprising once one realizes that $c'$, being the universal constant of
the Casimir effect, measures the number of physical degrees of freedom
of the system, and it is quite reasonable to expect that non-anomalous
interaction terms do not alter this number.

\vskip.3cm
Another kind of interaction has also been considered, proportional to
the square of the extrinsic curvature of the world sheet
{}~\cite{{helf},{PO}}, which favors
more realistic smooth configurations of the flux tube at short distance.
 Even this interaction does not modify the asymptotic universal behaviour
(\ref{frasy}, \ref{c}) of the free bosonic string ~\cite{bra}
(see however~\cite{oy}).

\vskip.3cm
Unfortunately, all these modifications of the Nambu-Goto action make it
difficult
to calculate, even approximately, the free energy $F(R,L)$ for finite
values of $R$ and $L$, so that a direct comparison of eq. (\ref{polya})
with numerical data is very problematic.

\vskip.3cm
There is however, in our opinion, a serious objection
against the validity of all these bosonic strings, due to the fact that
in
such a case only one kind of boundary conditions can be considered,
namely
the fixed ones: the transverse displacements $X_i$ must vanish on the
boundaries. In such a case, there is no way to distinguish the two ends
of the string and the constraint (\ref{empty}) cannot be satisfied.
In other words, there is no selection rule
which prevents a bosonic
string coming from a quark source
reaching the same source again, while the colour flux tube emitted by a
quark in a confined phase can never go back to the source, of course.

\vskip.3cm
In order to have an effective string which behaves like a colour flux
tube, it is necessary to supplement the theory with some conserved
quantum number, which makes it possible to distinguish the two ends of
the string, as eq.
(\ref{empty}) and (\ref{z2}) require. One possibility, suggested by the
superstring, is to introduce new degrees of freedom, besides the
transverse displacements, which however modify the conformal anomaly
(\ref{c}) and are difficult to justify on general grounds.

\vskip.2cm
There is a simpler, more convincing way to modify the bosonic string
without changing $c$, based on the observation that the real flux tube
connecting a pair of quarks has a small, but not vanishing, thickness.
Then it is reasonable to expect that the quantum fluctuations are
formed not only by
the local deformations of the string, but also by topological excitations
characterized by the number of times the string wraps around the flux
tube. Expanding the
transverse displacements $X_i$ of the string around a classical solution
$X^{cl}_i$ , we can implement the winding modes of the string by
assuming the quantum fluctuations $x_i=X_i-X^{cl}_i$ compactified~
\cite{cremmer}
\eq
x_i\equiv x_i+2\pi n\,R_t~~~~~~~~~,
\en
where $R_t$ defines an effective radius of the flux tube. The conformal
anomaly (8) is not affected by these topological excitations, but the
spectrum of the allowed conformal operators of the quantum theory is
drastically modified and is a function of $R_t$.

\vskip.2cm
An interesting feature of this one-parameter family  of conformal
theories is that, in many cases, there is a free fermion in the spectrum.
 This is a common property of dynamical  systems which can sustain local
and topological modes. A free fermion in the spectrum accounts for
another property of the color flux tubes: they cannot self-overlap
freely, as also strong coupling expansions seem to require~\cite{david}.
No such constraints exist in the usual bosonic string.

\vskip.2cm
This suggests assuming that the effective string, at least at large
distances, is described by one of  these free fermion theories. The
boundary condition of this fermion is a function of $R_t$. We show now
that eq.(4) and (6) uniquely fix this boundary condition as well as
$R_t$.

Consider a free Dirac fermion $\psi(\varsigma,\tau)=\psi^1-i\psi^2$ on an
infinite strip of width $R$ with action
\eq
S=-\frac{i}{2}\int{\rm d}\tau\int_{-\frac{R}{2}}^{\frac{R}{2}}{\rm d}
\varsigma\,\bar{\psi}\not\partial\,\psi~~~~.\en
Equations of motion say that
\eq
\psi_\pm^a(\varsigma,\tau)=\psi_\pm^a(\tau\pm\varsigma)~~~~,
\en
where $\psi_+^a$ and $\psi_-^a$ are the up and down components of the
Majorana fermion $\psi^a$. Because of the finite width, in the variation
of $S$ arises also a boundary term
\eq
\psi_+^a\delta\psi_+^a-\psi_-^a\delta\psi_-^a~~,~~\varsigma=\pm\frac{R}{2}~,
\en
which implies $\psi(\varsigma,\tau)\equiv0$ unless one makes the rather
arbitrary assumption $\delta\psi_+=\pm\delta\psi_-$~.

\vskip.2cm
A better way to treat this problem is to add a boundary term $S
_B$~\cite{ademollo} to the bulk action (10) in order to compensate
the contribution (12). The most general conformal invariant, hermitian
term, has the form
\eq
S_B(\varsigma=\frac{R}{2})=\frac{i}{2}{\rm cos}(2\pi\nu)\int{\rm d}\tau
\,\psi_-^a\psi_+^a+\frac{i}{2}{\rm sin}(2\pi\nu)\int{\rm d}\tau\,
\epsilon_{ab}\psi^a_-\psi^b_+~~~,
\en
where $\epsilon_{12}=-\epsilon_{21}=1\,,\,\epsilon_{aa}=0$  and $\nu$ is a
not yet specified boundary phase. A similar expression holds for the
other side at $\varsigma=-\frac{R}{2}$ , with another phase $\nu'$, but it
is always possible to redefine the field $\psi$ such that
\eq
S_B(\varsigma=-\frac{R}{2})=\frac{i}{2}\int{\rm d}\tau\,
\psi_+^a\psi_-^a~~~~.
\en
When $\delta S_B$ is combined with eq.(11) and (12) we get
\eq
\psi_\pm(\varsigma+2R,\tau)={\rm e}^{\pm2\pi i\nu}\psi_\pm(\varsigma,\tau)~~~.
\en
Modular invariance implies rational values for $\nu$. These boundary
conditions uniquely fix the conformal spectrum of the theory. Comparison
with the spectrum of the bosonic formulation supplemented with the
compactification (9) yields
\eq \nu=\sqrt{\pi\,\sigma}R_t~~~~,
\en
which relates the conditions at the quark boundaries to the thickness of
the colour flux tube.

\vskip.2cm
We come now to the constraints (4) and (6). Comparing eq.(13) and eq.(14)
 shows that the term proportional to ${\rm cos}(2\pi\nu)$ at $\varsigma=
\frac{R}{2}$ is compatible with that at $\varsigma=-\frac{R}{2}$ ,
therefore the amplitude for a direct open-closed string transition is
forbidden, as required by eq.(4), only if ${\rm cos}(2\pi\nu)=0$, i. e.
\eq
\nu=\frac{1}{4}~~~~~.
\en
In turn, this fixes the universal coefficient of the asymptotic
behaviour (3)
\eq
c=D-2~~~~,~~~~\frac{c}{24}-h_{\alpha\beta}=\frac{D-2}{96}~~.
\en
To see that also eq.(6) is fulfilled, consider the reparametrization
$\varsigma\rightarrow-\varsigma$~,which is a symmetry of the bulk action and
transposes the two sides of the strip. We may implement this symmetry
with the following field transformation
\eq
\psi^a_+\rightarrow\psi^a_-~~~\psi^1_-\rightarrow\psi^2_+~~~ \psi^2_-
\rightarrow-\psi^1_+~~,
\en
which exchanges also the two boundary terms (13) and (14), as eq.(6)
requires.

\vskip.2cm
In conclusion, a universal string picture describing the large-distance
behaviour of any gauge theory in the confining phase emerges rather
naturally. We have essentially utilized only the following three obvious
assumptions

\noindent
i) each quark is connected to only one flux tube,

\noindent
ii) colour flux tubes have a non-vanishing thickness and

\noindent
iii) they cannot overlap.

{}From these assumptions , conformal field theory tells us that at large
distances, where the troubles with longitudinal modes are disappearing
\cite{olesen}, the only consistent, effective string is a free fermion
model with boundary phase $\nu=\frac{1}{4}$. This is exactly the string
picture which was proposed some time ago~\cite{noi1} on purely
phenomenological grounds, where the boundary phase $\nu=\frac{1}{4}$
were determined only by a best fit to the numerical data on the
expectation values of Wilson loops. This picture as been proven~\cite
{{noi2},{noi3}} to be quite good for all those $3D$ and $4D$ gauge
systems where accurate numerical data are available.

\vskip.2cm
In this paper we
have found that this boundary phase is related to the size of the flux
tube through eq.(16), yielding  $\sqrt{\sigma}R_t\simeq 0.14~$.
This gives a lowest bound on the range of validity of this string
picture, for the minimal length $R_{c}$ of the string must be larger
than its thickness. Numerical data actually give $\sqrt{\sigma}R_c\simeq
0.3\div 0.4$~\cite{noi3}.

\vskip.2cm
In the free string approximation, it is also possible
to evaluate the range of validity of
the two asymptotic expansions (2) and (3), because now the free energy
can be evaluated exactly for arbitrary $R$ and $L$. Putting
$F(R,L)=\sigma RL+q(R,L)$~, where $q(R,L)$ is the quantum contribution
and is only a function of the scaling variable $z=\frac{2R}{L}\,$,
we have
\eq
q(R,L)=
-(D-2)\sum_{m=0}^3\log\frac{\vartheta\spinst{m/4}
{1/4}(0|\tau)}{ \eta(\tau)}
\en
where $\tau=i/z$~, $\vartheta\spinst{\alpha}{\beta}(0|\tau)$ is the
Jacobi theta
function with characteristic $\spinst{\alpha}{\beta}$ and $\eta(\tau)$
is the Dedekind eta function. This function is plotted in fig.1 with
other kinds of free string. It shows that the two asymptotic regions (2)
and (3) , where a universal signal of the string could be observed,
correspond respectively to
$z > 5.$ and  $z < 0.3$ , so they are rather far from the range explored by
the ordinary numerical simulations \cite{engels}, except for
ref.\cite{berg} ( $0.5\leq z\leq 16$) where eq.(8) is found not
inconsistent with the data, and this explains why no clear signal of a
string has yet been reported in Polyakov loop analysis.

\vskip 1cm
One of us (M.C.) would like to thank L. Magnea, P. Lepage, D. Lewellen
and  S. Tye for useful discussions during his permanence in Cornell.


\vskip .6 cm
\centerline{\sl Figure Captions}
\vskip .3 cm
\begin{description}

\item{Fig.1}
Quantum string contributions, in the cylinder  geometry, for various
values of the ratio $z=\frac{2R}{L}$. The continuous line is the fermion
string with $\nu=\frac{1}{4}$, the dotted line is the bosonic string and
the dashed line is the $c=\frac{1}{2}$ critical Ising model.
\end{description}

\vfil
\end{document}